\documentclass[aps,prl,twocolumn,amsmath,amssymb,showpacs]{revtex4-1}

\usepackage[ansinew]{inputenc}
\usepackage{graphicx}
\usepackage{textcomp}
\usepackage{xcolor}
\usepackage{amsmath}
\usepackage{amssymb}
\usepackage{siunitx}
\usepackage{hyperref}

\begin{document}

\title{Purcell enhancement of single-photon emitters in silicon}
\author{Andreas Gritsch}
\author{Alexander Ulanowski}
\author{Andreas Reiserer}
\email{andreas.reiserer@tum.de}

\affiliation{Max-Planck-Institute of Quantum Optics, Quantum Networks Group, Hans-Kopfermann-Stra{\ss}e 1, 85748 Garching, Germany}
\affiliation{Technical University of Munich, TUM School of Natural Sciences, Physics Department and Munich Center for Quantum Science and Technology (MCQST), James-Franck-Stra{\ss}e 1, 85748 Garching, Germany}

\begin{abstract}
Individual spins that are coupled to telecommunication photons offer unique promise for distributed quantum information processing once a coherent and efficient spin-photon interface can be fabricated at scale. We implement such an interface by integrating erbium dopants into a nanophotonic silicon resonator. We achieve spin-resolved excitation of individual emitters with $< \SI{0.1}{\giga\hertz}$ spectral diffusion linewidth. Upon resonant driving, we observe optical Rabi oscillations and single-photon emission with a 78-fold Purcell enhancement. Our results establish a promising new platform for quantum networks.
\end{abstract}

\maketitle

\section{Introduction}

The spins of individual photon emitters in solids offer unique promise for quantum technology \cite{awschalom_quantum_2018}, as they can combine excellent coherence with scalable manufacturing. To unlock their full potential, it is mandatory to establish a coherent and efficient spin-photon interface. This can be achieved by integrating the emitters into optical resonators \cite{reiserer_colloquium_2022}. After pioneering experiments with trapped atoms \cite{reiserer_colloquium_2022} and quantum dots \cite{lodahl_interfacing_2015}, color centers in diamond have emerged as a prominent platform \cite{ruf_quantum_2021} that combines efficient spin-photon coupling with long-term memory and thus enables advanced quantum networking protocols \cite{bhaskar_experimental_2020, hermans_qubit_2022}. However, manufacturing diamond-based devices at scale is an outstanding challenge and so far requires sophisticated techniques for heterogeneous integration \cite{wan_large-scale_2020}. Therefore, the use of other host materials for single emitters, such as silicon carbide \cite{lukin_4h-silicon-carbide--insulator_2019, babin_fabrication_2022} and silicon, seems advantageous. 

In the latter, nanofabrication has reached a unique level of maturity. This enables optical resonators with small mode volume and exceptional quality factors, exceeding $10^7$ \cite{asano_photonic_2017}, which can be fabricated with tight tolerances on 300-mm-wafers \cite{ashida_ultrahigh-q_2017}. In addition, isotopically purified silicon, which can host spin qubits with exceptional coherence \cite{saeedi_room-temperature_2013} and narrow inhomogeneous broadening \cite{chartrand_highly_2018}, can be obtained at scale by epitaxial growth on photonic silicon-on-insulator samples \cite{liu_28silicon--insulator_2022}. Finally, not only nanofabrication but also the doping of silicon for information processing devices is well-established, which makes it an ideal platform for programmable classical \cite{bogaerts_programmable_2020} and quantum \cite{pelucchi_potential_2021} photonics circuits.

To integrate emitters at telecommunication wavelengths into silicon, two approaches have been investigated. First, a broad diversity of color centers with promising single-emitter properties has been explored \cite{redjem_single_2020, durand_broad_2021, hollenbach_wafer-scale_2022}. Among them, the T-center has enabled spin-resolved detection in a nanostructured material \cite{higginbottom_optical_2022}, and -- in parallel to our current work -- an increase of the fluorescence  \cite{lefaucher_cavity-enhanced_2023, saggio_cavity-enhanced_2023} and a reduction of the lifetime \cite{redjem_all-silicon_2023} of G-centers has been observed upon integration into nanophotonic resonators. The second approach to embed photon emitters into silicon -- followed here -- is the use of single erbium dopants \cite{yin_optical_2013}. Their optical transitions fall within the main band of optical communication, where loss in optical fibers is minimal. In addition, they can exhibit narrow inhomogeneous linewidth \cite{weiss_erbium_2021, berkman_observing_2023} and long optical coherence at a temperature of $\SI{4}{\kelvin}$ even in nanophotonic waveguides \cite{gritsch_narrow_2022}.

In spite of the promise of single erbium dopants in silicon for quantum networking, their integration into optical resonators has not been demonstrated. In contrast, resonator-integration of rare-earth dopants in yttrium orthosilicate and yttrium orthovanadate has enabled efficient \cite{zhong_optically_2018, dibos_atomic_2018} and coherent \cite{ulanowski_spectral_2022} single photon generation, single-shot spin detection \cite{raha_optical_2020, kindem_control_2020}, and frequency-multiplexed control \cite{chen_parallel_2020} of coherent emitters with ultra-narrow spectral diffusion linewidth \cite{ulanowski_spectral_2022, ourari_indistinguishable_2023}. All of these techniques require resonators with strong Purcell-enhancement \cite{reiserer_colloquium_2022}. In this work, we show that this can also be achieved with nanophotonic silicon resonators, which offer unique prospects for scalable manufacturing.

\begin{figure*}[htb]
\includegraphics{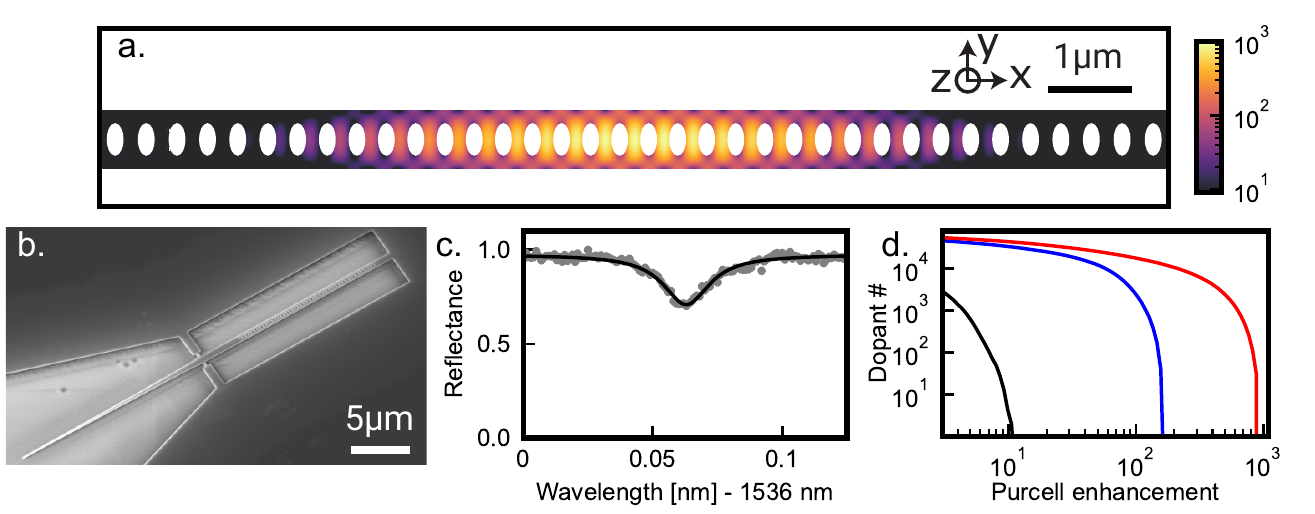}
\caption{   \label{fig:Figure1}
\textbf{Photonic crystal resonator.} a) Geometry and simulated maximum Purcell enhancement (colors / grey) of the fundamental resonant TE-like mode. b) Scanning electron microscope image of a typical device. c) Reflection spectrum after gas condensation tuning, normalized to its maximum. A dip is observed at the desired resonance wavelength. d) Simulated number of dopants whose Purcell enhancement exceeds the value given on the x-axis. A quick drop is observed when the maximum Purcell factor is approached. This also depends on the emitter orientation - strongest enhancement is observed for emitters with a dipole along $y$ (top red), weaker along $x$ (blue) and almost no enhancement along $z$ (bottom black). The coordinate system is defined in panel a.
}
\end{figure*}

\section{Erbium-doped nanophotonic resonator}

In our experiment, we integrate erbium at a recently discovered site B \cite{gritsch_narrow_2022} that exhibits a lifetime that is predominantly radiative and shorter than in any other known host material ($\SI{186}{\micro\second}$). It further combines narrow inhomogeneous ($\sim \SI{0.5}{\giga\hertz}$) and homogeneous ($\leq \SI{0.01}{\mega\hertz}$) linewidths with a large splitting of the crystal field levels. The sample fabrication is outlined in \cite{gritsch_narrow_2022}. In short, we start with a chip obtained by chemical vapor deposition on a thin silicon-on-insulator wafer. It is implanted at $\sim \SI{800}{\kelvin}$ with erbium to a peak concentration of $0.2 \times 10^{18} \si{\centi\meter}^{-3}$ and then diced in $10 \times 10 \si{\milli\meter}^2$ pieces. Nanophotonic structures are then fabricated using electron-beam lithography and reactive ion etching in fluorine chemistry. At the used implantation dose, many emitters are present in mm-long waveguides. Instead, in this work, we study a few-$\si{\micro\meter}$ short one-dimensional photonic crystal nanobeam cavity, which contains only a couple of emitters. The resonator is efficiently coupled to a feed waveguide \cite{quan_photonic_2010} and from there to a single-mode optical fiber via an adiabatically tapered transition \cite{tiecke_efficient_2015}. 

The geometry of the used resonator is shown in Fig. \ref{fig:Figure1}a,  and in the scanning electron microscope image in panel b. To design it, we perform finite-difference-time-domain simulations using the free and open-source software MEEP \cite{oskooi_meep_2010}. We start from a waveguide of $\SI{0.7}{\micro\meter}$ width and $\SI{0.19}{\micro\meter}$ thickness. We calculate the two lowest-lying TE-bands for a photonic crystal unit cell with elliptical holes of $\SI{325}{\nano\meter}$ major and $\SI{150}{\nano\meter}$ minor axis length. At the fixed target wavelength of the resonator, $\SI{1536.06}{\nano\meter}$, we then choose the spacing between neighboring holes such that the mirror strength --- calculated from the band edges --- is linearized \cite{quan_photonic_2010}, which minimizes scattering loss. We use a single-sided device, which improves the outcoupling to the feed waveguide, with 20 holes at the outcoupler and 25 on the other side. The simulated mode volume is $V = 1.45 (\lambda / n)^3$, where $\lambda$ is the wavelength and $n$ the refractive index. Measuring the linewidth of the resonator in reflection at cryogenic temperature (Fig. \ref{fig:Figure1}c), $\SI{2.7}{\giga\hertz}$, we extract $Q=0.73(3)\times 10^5$, significantly lower than the simulated value of $Q>10^5$ because of fabrication imperfections that may be largely eliminated in future devices by an improved lithography and careful optimization of the etching \cite{asano_photonic_2017}. At the used erbium concentration, we observe no significant deterioration of Q in implanted as compared to unimplanted samples.

For a resonant emitter at the cavity field maximum whose transition dipole moment is perfectly aligned with the cavity, and in the absence of nonradiative decay, at the mentioned parameters we expect a $P=3.8 \times 10^3$-fold Purcell enhancement, defined as the reduction of the emitter lifetime as compared to bulk silicon \cite{reiserer_colloquium_2022}. This reduces to $P=0.88(19) \times 10^3$ when considering the $23(5)\,\%$ branching ratio of the erbium transitions in silicon \cite{gritsch_narrow_2022}. The dependence of the maximally expected lifetime reduction on the emitter position in the center plane of the cavity is shown in Fig. \ref{fig:Figure1}a (colors). Using the simulated three-dimensional field distribution and implantation depth profile for the implantation parameters mentioned above, the maximally expected number of dopants with a Purcell enhancement larger than a certain value is shown in Fig. \ref{fig:Figure1}d, depending on the orientation of the optical dipole. The experimentally observed number of dopants will be much lower because only a small fraction of the implanted dopants is integrated at site B \cite{gritsch_narrow_2022}.

\begin{figure*}[tb]
\includegraphics{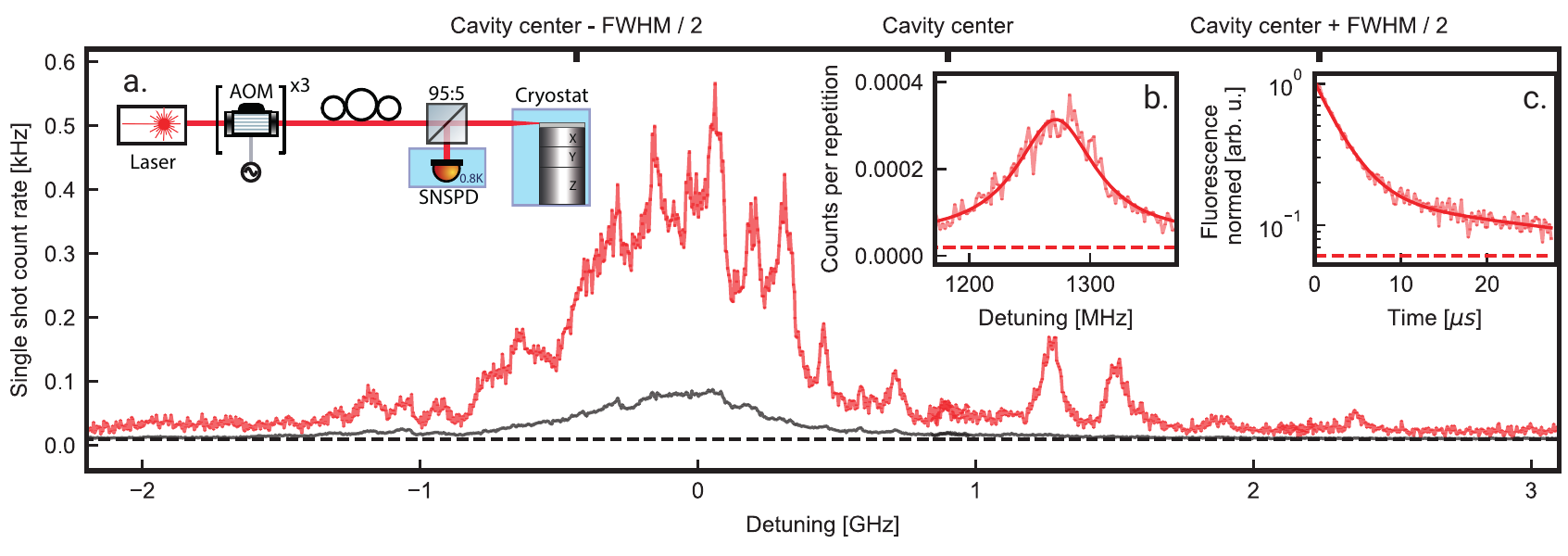}
\caption{     \label{fig:FluorescenceSpectrum}
\textbf{Single dopant spectroscopy.} Inset a: Experimental setup. A frequency-stabilized laser is switched by fiber-coupled acousto-optic modulators (AOM), and its polarization is controlled by fiber-optical paddle controllers. A 95:5 beam splitter is used to guide the light emitted from the sample in the cryostat to single-photon detectors (SNSPD). Main graph: Fluorescence spectroscopy. The frequency of the excitation laser pulses is scanned while the cavity frequency is fixed (top axis) at $\sim {\SI{1}{\giga\hertz}}$ detuning (bottom axis) from the center of the inhomogeneous line of the erbium dopants. On long timescales (dark grey, $\SI{10}{\micro\second} - \SI{48}{\micro\second}$ after the excitation pulse), the signal is dominated by dopants in the feed waveguide. On shorter times (light red, $< \SI{1}{\micro\second}$ after the pulse), several peaks that originate from single dopants are observed. Inset b: Zoom into a region of the spectrum with a single, well-resolved emitter. Inset c: Temporal decay of the fluorescence measured at the center of the peak in b.}
\end{figure*}

\section{Single-dopant spectroscopy}
Common fabrication fluctuations lead to shifts in the resonance wavelength of the cavity. This could be avoided by pre-implantation laser processing of the silicon chips \cite{lee_local_2009}. To avoid the added effort while still having a resonant cavity on every sample, we instead fabricate many resonators on the same chip, with a linear change of the design wavelength in $\SI{1.5}{\nano\meter}$ steps. After mounting the sample in a closed-cycle cryostat in He exchange gas with a variable temperature $T \geq \SI{1.7}{\kelvin}$, we select the resonator that is closest to the erbium transition wavelength. We then tune it on resonance by temporarily opening a valve to condense a thin film of argon ice to the sample surface \cite{mosor_scanning_2005}. On the first sample studied in this work, this required a shift of approximately \SI{0.35}{\nano\meter}.

As the lifetime of the spin levels of erbium emitters in silicon is temperature dependent, see \cite{gritsch_narrow_2022}, we choose optimal temperature settings for the different measurements in this manuscript. We start with pulsed resonant spectroscopy at $T=\SI{14}{\kelvin}$, which is cold enough to avoid broadening of the erbium transitions, but warm enough to avoid pumping to long-lived dark spin states \cite{gritsch_narrow_2022} when the Zeeman splitting exceeds the pulse bandwidth. We excite the dopants every $\SI{50}{\micro\second}$ with laser pulses of $\SI{0.4}{\micro\second}$ duration, which gives a spectral resolution of $\lesssim \SI{1}{\mega\hertz}$. The used setup is sketched in Fig. \ref{fig:FluorescenceSpectrum}a. We then scan the laser wavelength while the cavity frequency remains unchanged at $\SI{0.9}{\giga\hertz}$ detuning from the center of the inhomogeneous line. Fig. \ref{fig:FluorescenceSpectrum} (main panel) shows the fluorescence in a time interval from $\SI{10}{\micro\second}$ to $\SI{48}{\micro\second}$ after the laser is turned off (dark gray). We observe a broad background from off-resonant dopants, as characterized earlier \cite{gritsch_narrow_2022}, which is slightly larger than the detector dark counts (dashed line). In addition, the spectrum exhibits a broad peak that is attributed to erbium dopants in the $\SI{0.05}{\milli\meter}$ long feed waveguide, as not only the resonator, but the whole chip is implanted with erbium. However, when restricting the analysis to the first $\SI{1}{\micro\second}$ after the pulse is turned off (red), pronounced peaks appear. Similar to related works in other host crystals \cite{dibos_atomic_2018, ulanowski_spectral_2022}, these features are attributed to individual dopants that are located in the resonator field and thus decay on a faster timescale. They exhibit different emission frequencies because of local strain. At the center of the inhomogeneous line, the peaks reside on a considerable background, whereas they are well-isolated at larger detunings where the density of emitters in the feed waveguide is sufficiently reduced. Fitting the peaks, we find spectral diffusion linewidths between $0.01$ and $\SI{0.1}{\giga\hertz}$. The large variation may indicate that the source of broadening is local to the respective emitters. We speculate that it is caused by fluctuating charge traps, as commonly observed in semiconductors \cite{kuhlmann_charge_2013}, and thus depends on the distance of the dopant to proximal interfaces. When comparing the number of peaks to Fig. \ref{fig:Figure1}d and assuming that only dopants with a lifetime reduction factor $P \geq 50$ are observed because of the short detection time window, we estimate that the fraction of dopants that is integrated at site B is below $0.1\,\%$. This is comparable to our earlier work \cite{gritsch_narrow_2022}.

\begin{figure*}[htb]
\includegraphics{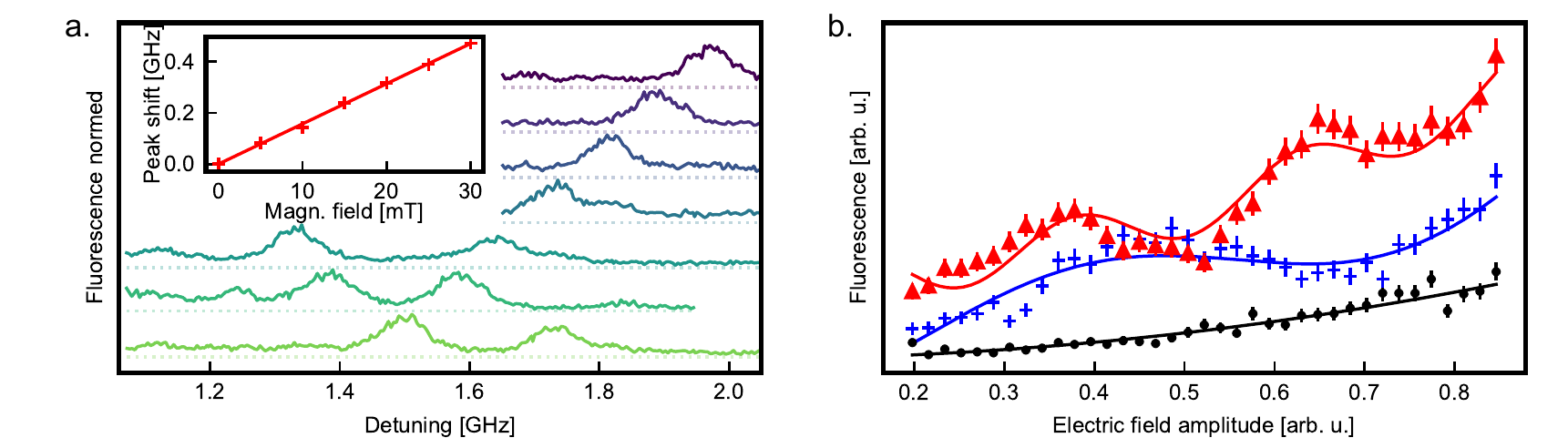}
\caption{     \label{fig:Rabi}
\textbf{a) Magnetic field dependence.} Increasing the magnetic field in steps of $\SI{5}{\milli\tesla}$ (bottom to top), the Zeeman states of the ground- and excited state of a single dopant at $\SI{1.5}{\giga\hertz}$ detuning split, which facilitates spin-resolved optical excitation. \textbf{ b) Optical Rabi oscillations}. The emission probability exhibits a Rabi oscillation when increasing the electric field amplitude of the excitation pulse, which proves coherence of the emitters during the pulses. At different polarization (colors, markers), the excitation efficiency is reduced and the period of the oscillation is increased. 
}
\end{figure*}

Fig. \ref{fig:FluorescenceSpectrum}b shows a zoom into the spectrum, in which a single, well-isolated dopant is found that exhibits a Lorentzian spectral-diffusion linewidth of $\SI{79(5)}{\mega\hertz}$. The value is of the same order of magnitude as that observed with erbium in other hosts and in the proximity of nanostructures \cite{chen_parallel_2020}, with the exception of Erbium in inversion-symmetric sites (e.g. $\SI{0.2}{\mega\hertz}$ in \cite{ourari_indistinguishable_2023}). It is about 20-fold lower than that of other emitters in nanophotonic silicon structures \cite{higginbottom_optical_2022, prabhu_individually_2023}. Still, it is much larger than the lifetime-limited linewidth of $\SI{0.0521(2)}{\mega\hertz}$, which we determine from the shorter timescale of a bi-exponential fit to the fluorescence decay (Fig. \ref{fig:FluorescenceSpectrum}c), where the longer timescale corresponds to weakly-coupled dopants in the cavity and feed waveguide. The obtained lifetime of $\SI{3.12(1)}{\micro\second}$ corresponds to a $P=59.7(2)$-fold Purcell enhancement compared to the bulk decay \cite{gritsch_narrow_2022}. The obtained $P$ value exceeds that of many previous works on quantum dots \cite{lodahl_interfacing_2015} and diamond nanophotonics \cite{bhaskar_experimental_2020, ruf_quantum_2021} and is comparable to that obtained with erbium in macroscopic Fabry-Perot resonators \cite{merkel_coherent_2020, ulanowski_spectral_2022}. It is seven times smaller than that obtained with erbium in other hosts in the proximity of silicon nanophotonic resonators \cite{dibos_atomic_2018}. However, because of the faster radiative decay rate in silicon, we still observe a shorter lifetime, which has a higher practical relevance than the enhancement factor. In our experiment, the latter is still 15(3)-fold lower than the theoretical maximum. This is not unexpected, since only emitters at the field maximum and with a matching dipole exhibit the maximum enhancement, c.f. Fig. \ref{fig:Figure1}d. By improving the precision of the gas condensation tuning (e.g. with an automated valve), using a superconducting vector magnet to determine and align the optical dipole orientation \cite{raha_optical_2020} to the cavity field, and by only implanting the cavity at the mode center, we expect that one can approach the theoretical Purcell enhancement value in future experiments and at the same time eliminate the background of weakly-coupled emitters.

\section{Observation of single spins}

The observation of individual dopants is a promising first step towards quantum information processing. To this end, one further requires spin-resolved optical transitions and coherent microwave driving, as observed previously with erbium dopants in other host materials \cite{raha_optical_2020, cova_farina_coherent_2021}, in which a magnetic field lifts the degeneracy of the effective spin-$1/2$ Zeeman levels. To investigate this in silicon, we apply a magnetic field along the (100) axis and again perform a pulsed fluorescence measurement. When the Zeeman splitting exceeds the pulse bandwidth, the signal disappears at low temperature as the dopants are quickly pumped to the long-lived other spin state that is off-resonant. Thus, to observe the spin splitting we again measure at $T=\SI{14}{\kelvin}$, which reduces the spin lifetime below that of the optically excited state \cite{gritsch_narrow_2022}. For the prominent peak in Fig. \ref{fig:FluorescenceSpectrum}b, a clear splitting is observed, see Fig. \ref{fig:Rabi}a, as the effective $g$-factor of the ground- and excited state of erbium differs in sites with low symmetry. From Lorentzian fits (data points in the inset), we determine $\Delta g = \SI{31.3(6)}{\giga\hertz\per\tesla}$ for this dopant. Applying the magnetic field along different directions will allow for extracting the full $g$ tensor in future experiments. Furthermore, this may facilitate single-shot spin readout by frequency-selective enhancement of one transition \cite{ourari_indistinguishable_2023} once better outcoupling is achieved.

\section{Rabi oscillations}

\begin{figure*}[tb]
\includegraphics{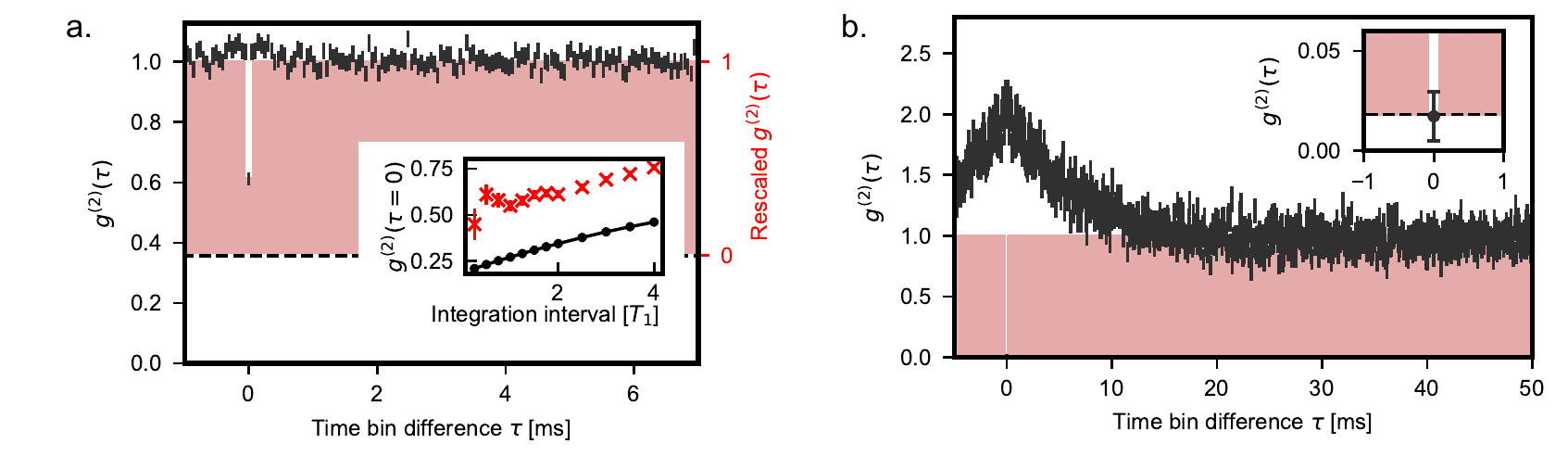}
\caption{     \label{fig:G2}
\textbf{Pulsed autocorrelation measurement.} After exciting a frequency-resolved single dopant in the cavity, the emitted light exhibits antibunching. a) In the sample with high implantation dose, a considerable background from dopants in the feed waveguide leads to a significant value at zero time delay, even after rescaling (right axis, red) to account for accidental correlations from detector dark counts (dashed line). Inset: The autocorrelation at zero delay $\tau=0$ for different integration time windows $T$ reveals the expected reduction of the dark count contribution (black). b) On a second sample with a $\sim 29$-fold lower implantation dose, the contribution of ions in the feed waveguide is reduced. In addition, improved outcoupling leads to a smaller dark-count contribution (see inset). This results in almost perfect antibunching at zero delay. Around that value, we observe bunching caused by the pumping into another ground-state spin level in the applied magnetic field of \SI{0.55}{\tesla}.
}
\end{figure*}

To use single dopants for quantum networking, it is required that the emitted light is coherent \cite{reiserer_colloquium_2022}. To investigate this, we turn off the superconducting magnet and excite the sample using pulses of $\SI{15}{\nano\second}$ duration. We then scan their polarization and intensity at $T=\SI{4}{\kelvin}$, such that the coherence is not limited by $T$ \cite{gritsch_narrow_2022}. Using short, high-bandwidth pulses in this measurement ensures that transitions from both Zeeman states are driven even in case a small residual magnetic field is present. The cavity-coupled emitters are only excited when the input polarization matches that of the resonator mode. As can be seen in Fig. \ref{fig:Rabi}a, in this case we obtain clear signatures of coherent oscillations on a quadratically rising background (red triangles and fit) which may be expected for the dopants in the feed waveguide. When reducing the polarization matching (blue crosses to black dots), the period is increased, and the overall excitation probability is reduced. The observation of Rabi oscillations upon varying the field amplitude proves that the emitters are coherently excited, and that the strong pulses do not induce instantaneous diffusion on the investigated timescale via two-photon or background-dopant absorption.

\section{Single photon emission}

The spectral properties, magnetic field splitting and Rabi oscillations all suggest the observation of single emitters in the resonator. To obtain further evidence, we use the emitters to generate single photons at telecommunication wavelength. To this end, we change $T$ to $\SI{10}{\kelvin}$ in order to avoid spin pumping while still achieving a narrow homogeneous linewidth, down to $\SI{0.2}{\mega\hertz}$ \cite{gritsch_narrow_2022}. This would be sufficiently narrow to enable lifetime-limited photon emission at the achieved Purcell enhancement. We then tune the laser to the emission of a well-resolved single dopant (that shown in Fig. \ref{fig:FluorescenceSpectrum}b).

We excite the system every $\SI{50}{\micro\second}$ with square laser pulses of $\SI{25}{\nano\second}$ and first determine the probability to detect a photon after an excitation pulse, which is $0.1\,\%$. The moderate value is explained by the finite efficiency of the available detector ($40\,\%$), fiber-coupling ($\sim 10\,\%$) and cavity outcoupling ($\sim 10\,\%$), as well as a finite emitter excitation probability, since the bandwidth of the excitation pulses, $\approx \SI{18}{\mega\hertz}$, did not exceed the emitter spectral diffusion linewidth. The efficiency can thus be enhanced by an improved pulse generation setup, by further optimization of the cavity design \cite{knall_efficient_2022} and by using underetched silicon waveguide tapers for efficient coupling, as we will show below.

Before, we obtain the autocorrelation function of the emitted light. To this end, we record all photon detection events registered by a single detector, which has a dead time of $< \SI{50}{\nano\second}$ that is negligible compared to the emission timescale such that a Hanbury Brown and Twiss setup is not required.  In post-processing, we then analyze the emission detected after the excitation pulse within a time bin of $T=\SI{14}{\micro\second}$ duration. This corresponds to twice the emitter lifetime $T_1$, as the Purcell enhancement is reduced in this measurement because of an undesired change of the resonator frequency after thermal cycling (compared to that shown in Fig. \ref{fig:FluorescenceSpectrum}). In the obtained autocorrelation function, we observe clear antibunching, see Fig. \ref{fig:G2}a (black data). Owing to the limited efficiency and long integration time, $g^{\left(2\right)}\left(\tau=0\right)$ has a significant contribution from accidental correlations with detector dark counts. These can be measured independently in the absence of laser excitation pulses (dashed line) and the axis can then be rescaled to obtain the values that would be obtained with perfect detectors \cite{becher_nonclassical_2001} (red area, right axis). With this correction, we find a value of $g^{\left(2\right)}\left(\tau=0\right)=0.39(5)$ which already proves that we can spectrally address individual emitters in the same, subwavelength volume, paving the way for frequency-multiplexed control \cite{chen_parallel_2020, ulanowski_spectral_2022}. Still, the value in the zero-delay time-bin is limited by the previously mentioned background of uncoupled dopants in the waveguide. This could be avoided by spatially-selective implantation \cite{hollenbach_wafer-scale_2022} of erbium. Instead, we perform measurements on a second device, made from CZ silicon \cite{gritsch_narrow_2022} that is homogeneously implanted with $1\times 10^{11}$ ions per $\si{\centi\meter^2}$, a $\sim 29$-fold lower dose. Thus, in this device the contribution of emitters in the coupling waveguide is negligible, and only one or few dopants will be found in a typical cavity.

To further enhance the waveguide-to-fiber coupling efficiency to $\gtrsim 50\,\%$, we removed the buried oxide layer underneath the structure by etching in HF, which also slightly improves the $Q$ factor \cite{asano_photonic_2017}. After this, the resonator was close to being critically coupled with $Q=1.08(17)\times10^5$, enhancing the single photon detection probability per attempt by a factor of 12 compared to the previous device. Instead of using Argon gas, we tune the emitters precisely on resonance by applying a magnetic field of $\SI{0.55}{ \tesla}$. We find only one well-coupled emitter in this device, with a Purcell enhancement factor of $P=78.4(4)$ and a spectral diffusion linewidth of $\SI{70(13)}{\mega\hertz}$. Both values are similar to those obtained in the first device. The improved outcoupling, however, leads to a much smaller contribution of dark counts to the measurement, as can be seen in in Fig. \ref{fig:G2}b. Because of the strongly reduced background from dopants in the feed waveguide, we now obtain $g^{\left(2\right)}\left(\tau=0\right) = 0.017(12)$, which is in agreement with the expectation from the independently measured detector dark counts (dashed line in the inset).

In addition to the antibunching at zero delay, we also observe bunching on a timescale of a few $\si{\milli\second}$ in this experiment. The reason is that we applied a magnetic field and set the cryostat temperature to $\SI{4}{\kelvin}$, where the spin lifetime exceeds the optical lifetime \cite{gritsch_narrow_2022}. Thus, application of the excitation laser leads to a pumping to the other Zeeman state, which is off-resonant with the laser and resonator. As a result, the probability to emit a second photon after detecting a first one decreases with the number of applied laser pulses. As expected, the bunching thus disappears when the magnetic field is turned off and the frequency shift of the emitter is compensated by resonator tuning.

\section{Summary and Outlook}
In summary, we have demonstrated that single erbium dopants in silicon can be resolved and that their emission can be enhanced using a nanophotonic resonator. They furthermore exhibit spin-selective optical transitions that allow for the generation of single photons in the main band of optical communication, where loss in optical fibers is minimal. This offers great promise for the implementation of quantum networks over large distances. Remarkably, from the achieved 78-fold enhancement of the radiative transition and from the $<\SI{10}{\kilo\hertz}$ homogeneous linewidth measured with erbium ensembles in nanophotonic waveguides \cite{gritsch_narrow_2022}, one expects lifetime-limited coherence, which is a key enabler for remote entanglement via photon interference \cite{reiserer_colloquium_2022}. To this end, it will be required to know the precise emission frequency before each entanglement attempt. This may be achieved via fast measurements and feedback, as pioneered with emitters in diamond \cite{bernien_heralded_2013}. Alternatively, the spectral diffusion linewidth may be further reduced by using surface terminations, integrating the emitters into PIN-junctions \cite{anderson_electrical_2019} that can be compatible with high Q silicon resonators \cite{xu_ultrashallow_2021, nakadai_electrically_2022}, or by using cavities that facilitate a larger distance to interfaces at comparable Purcell enhancement \cite{ulanowski_spectral_2022}. In addition, the lifetime-limited linewidth may be further increased in resonators with larger quality factor \cite{asano_photonic_2017}. Taken together, these advances would establish erbium dopants in silicon as a prime candidate for large-scale quantum computing and communication networks.

\section*{Funding}
This project received funding from the European Research Council (ERC) under the European Union's Horizon 2020 research and innovation programme (grant agreement No 757772), from the Deutsche Forschungsgemeinschaft (DFG, German Research Foundation) under Germany's Excellence Strategy - EXC-2111 - 390814868 and via the project RE 3967/1, and from the German Federal Ministry of Education and Research (BMBF) via the grant agreements No 13N15907 and 16KISQ046.

\section*{Acknowledgements}
We acknowledge the technical contribution of Jakob Pforr and Nilesh Goel in the fabrication of the nanophotonic devices.

\bibliography{bibliography.bib}
\end{document}